\documentclass[12pt,notoc]{JHEP3}

\usepackage{amsmath,amssymb,euscript,array,cite,mathrsfs}

\def\newf{{\cal F}}

\usepackage{epsfig}

\def\rc{\check R}

\def\C{\Gamma}

\newcommand{\Tr}{\operatorname{Tr}}
\newcommand{\IM}{\operatorname{Im}}
\def\Ba{{\boldsymbol a}}
\def\Bb{{\boldsymbol b}}
\def\Bc{{\boldsymbol c}}
\def\Bd{{\boldsymbol d}}
\def\Bz{{\boldsymbol e}}
\def\Bmu{{\boldsymbol \mu}}
\def\Bnu{{\boldsymbol \nu}}
\def\Brho{{\boldsymbol \rho}}

\def\Bomega{{\boldsymbol \omega}}

\def\B0{{\boldsymbol 0}}
\def\BF{{\boldsymbol F}}
\def\BK{{\boldsymbol K}}
\def\BX{{\boldsymbol X}}
\def\BZ{{\boldsymbol Z}}

\def\BOmega{{\boldsymbol\Omega}}
\def\Bvarpi{{\boldsymbol\varpi}}
\def\Bpi{{\boldsymbol\pi}}
\def\CP{{\mathbb C}P}

\def\Tr{{\rm Tr}}

\def\det{{\rm det}}

\def\C{{\mathbb C}}

\def\R{{\mathbb R}}

\newcommand{\BH}{\boldsymbol{H}}
\newcommand{\Bone}{\boldsymbol{1}}
\newcommand{\Be}{\boldsymbol{e}}

\def\Dbarslash{\,\,{\raise.15ex\hbox{/}\mkern-12mu {\bar D}}}
\def\Dslash{\,\,{\raise.15ex\hbox{/}\mkern-12mu D}}
\def\delslash{\,\,{\raise.15ex\hbox{/}\mkern-9mu \partial}}
\def\delbarslash{\,\,{\raise.15ex\hbox{/}\mkern-9mu {\bar\partial}}}

\def\LAG{\mathscr{L}}

\newcommand{\MAT}[1]{\begin{pmatrix} #1\end{pmatrix}}
\newcommand{\ARR}[1]{\begin{matrix} #1\end{matrix}}
\newcommand{\EQ}[1]{\begin{equation}\begin{split} #1
\end{split}\end{equation}}

\newcommand{\SP}[1]{\begin{equation}\begin{split} #1
\end{split}\end{equation}}

\title{The Relativistic Avatars of Giant Magnons and their S-Matrix}
\author{Timothy J. Hollowood\\
Department of Physics,\\ University of Wales Swansea,\\
Swansea, SA2 8PP, UK.\\
E-mail: \email{t.hollowood@swansea.ac.uk}}
\author{and J.~Luis Miramontes\\
Departamento de F\'\i sica de Part\'\i culas and IGFAE,\\
Universidad
de Santiago de Compostela\\ 15782 Santiago de Compostela, Spain\\
E-mail: \email{jluis.miramontes@usc.es}}

\abstract{The motion of strings on symmetric space target spaces underlies
 the integrability of the AdS/CFT
  correspondence. Although these theories, whose excitations are giant 
magnons, are non-relativistic they are
  classically equivalent, via the Polhmeyer reduction, to a 
relativistic integrable field theory known as a 
symmetric space sine-Gordon theory. These theories can be formulated
as integrable deformations of gauged WZW models. In
  this work we consider the class of symmetric spaces $\CP^{n+1}$ and
  solve the corresponding generalized sine-Gordon theories at the quantum level
  by finding the exact spectrum of topological solitons, or kinks, and their
  S-matrix. The latter
  involves a trignometric solution of the Yang-Baxer equation which
  exhibits a quantum group symmetry with a tower of states that is
  bounded, unlike for magnons, as a result of the quantum  group deformation parameter $q$
  being a root of unity. We test the S-matrix by taking the semi-classical limit and comparing
with the time delays for the scattering of classical solitons. We
argue that the internal $\CP^{n-1}$ moduli space of collective coordinates of the
solitons in the classical theory can be interpreted as a $q$-deformed
fuzzy space in the quantum theory.
We analyse the $n=1$ case separately and provide a further test of the
S-matrix conjecture in this case by calculating the central charge of
the UV CFT using the thermodynamic Bethe Ansatz.}

\begin{document}

\section{Introduction}

One of the many remarkable features of the AdS/CFT correspondence is
the emergence of integrability. This is fortunate indeed, because it
promises a quantitative investigation of the conjectured duality. On the CFT side, integrability is manifested by the appearance of integrable spin chains whose Hamiltonians provide the spectrum of exact scaling dimensions $\Delta$~\cite{Zarembo:2004hp,Minahan:2006sk,Minahan:2008hf}. In the particular limit where $\Delta$ and a conserved $R$-charge $J$ become infinite, with the difference  $\Delta-J$ and the 't~Hooft coupling $\lambda$ held fixed, the string duals of the fundamental magnon excitations of those spin chains are lump-like solutions known as ``giant magnons'', which propagate on an infinite long string~\cite{Hofman:2006xt}.
Giant magnons describe the classical motion of (bosonic) strings on curved space-times of the form $\R_t\times {\mathfrak M}$, with ${\mathfrak M}=F/G$ a symmetric space. For example, the (original) giant magnon of $\text{AdS}_5/\text{CFT}_4$ and its dyonic generalization correspond to $S^n=SO(n+1)/SO(n)$ with $n=2$ and~$3$, respectively~\cite{Hofman:2006xt,DyonicGM}. In a similar way, the basic giant magnons of $\text{AdS}_4/\text{CFT}_3$ are associated to $\CP^n=SU(n+1)/U(n)$ with $n=1$ and~$2$~\cite{Gaiotto:2008cg,Grignani:2008is,Hollowood:2009sc,Abbott:2009um}.

The gauged-fixed worldsheet theory on $\R_t\times {\mathfrak M}$ is a
sigma model with target space ${\mathfrak M}$ subject to additional
constraints that preserve integrability, but break conformal and
relativistic invariance on the worldsheet. The gauge fixing conditions
are the Pohlmeyer constraints~\cite{Tseytlin:2003ii,Pohlmeyer:1975nb},
and the giant magnons are the solitons of the resulting constrained theory. 
Their spectrum and S-matrix has been completely
determined at the quantum
level~\cite{Staudacher:2004tk,BeisertSM,Arutyunov:2006yd,Ahn:2008aa}. The
S-matrix is complicated by the fact that, in comparison with the
``usual'' situation, the worldsheet theory is non-relativistic.  
It is remarkable, however, that there is a re-formulation of the sigma
model with Pohlmeyer constraints as a conventional massive integrable
field theory of a type that generalizes the sine-Gordon theory. These
relativistic field theories are known as the symmetric space
sine-Gordon (SSSG) theories \cite{Eichenherr:1979yw,Eichenherr:1979uk,D'Auria:1979tb,D'Auria:1980cx,D'Auria:1980xs,Zakharov:1973pp}. They can formulated at the Lagrangian
level as a gauged WZW model with an integrable deforming potential~\cite{Bakas:1995bm,Miramontes:2008wt}, which naturally leads to their description as perturbations of coset CFTs~\cite{CastroAlvaredo:2000kq}. The
equivalence between the gauged fixed worldsheet theory and the SSSG theory is a
classical equivalence in which 
the non-relativistic magnons map to a relativistic soliton ``avatar'' in the SSSG
theory. It does not seem possible that the equivalence can be
maintained at the quantum level, since the two descriptions have a
different Poisson structure~\cite{Poisson}. However, in the context of $\text{AdS}_5/\text{CFT}_4$, it has been argued by Grigoriev and Tseytlin~\cite{Grigoriev:2007bu} and by Mikhailov and Sch\"afer-Nameki~\cite{Mikhailov:2007xr} that quantum
equivalence may hold in the full theory with all the fermions
included. Then, the Lagrangian formulation of the SSSG theory would be the starting point to find a novel, manifestly
two-dimensional Lorentz invariant, formulation of the full
$\text{AdS}_5 \times S^5$ superstring theory that would be an alternative to the usual formulation in the light-cone gauge. This conjecture has already passed a number of tests~\cite{Tests}.
Nevertheless, the equivalence can only be properly judged once the SSSG theories have
been solved at the quantum level. To date, the knowledge of the SSSG
theories is extremely limited to the cases with no fermions and then only to
the SSSG theories
related to $S^2=SO(3)/SO(2)$ and $S^3=SO(4)/SO(3)$, since these are
the well-known sine-Gordon~\cite{Zamolodchikov:1978xm} and complex sine-Gordon~\cite{Dorey:1994mg}
theories, respectively.\footnote{Although they are not directly relevant in the context of the AdS/CFT correspondence, the homogeneous sine-Gordon theories provide another set of SSSG theories that have been solved at the quantum level~\cite{FernandezPousa:1996hi,Miramontes:1999hx,CastroAlvaredo:1999em,Dorey:2004qc}. They are Pohlmeyer reductions of the principal chiral model corresponding to a Lie group $G$, which can be formulated as the symmetric space $G\times G/G$~\cite{Miramontes:2008wt}.} The aim of the present work is to begin to fill
the gap in our knowledge by solving---in the sense of finding the spectrum
and S-matrix---the theories corresponding to the symmetric spaces
$\CP^{n+1}$. This is directly relevant to the AdS/CFT correspondence for $\text{AdS}_4\times\CP^3$; namely, $\text{AdS}_4/\text{CFT}_3$~\cite{NewDuality}. The extension to other symmetric spaces should now follow
by similar methods.

The plan of the paper is as follows. In section~\ref{PohlmeyerR} we review the Pohlmeyer reduction for the example $\CP^{n+1}$ focussing particularly on the algebraic approach that leads to the associated SSSG theory in a rather simple way. We explain
how the SSSG theory can be formulated at the Lagrangian level as a
gauged WZW model for $U(n+1)/U(n)$ with an integrable deforming potential. In section~\ref{Dressing},
we show that the non-relativistic magnons in the original sigma model can be constructed at the same time as the soliton avatar using the
dressing method. We spend some time explaining how the
magnons/solitons have a $\CP^{n-1}$ moduli space of internal collective
coordinates. We also show how the soliton avatar is a kink carrying a
topological charge.
In section~\ref{Smatrix} we present our conjecture for the exact
quantum S-matrix of the topological kinks of the deformed WZW
model. Section~\ref{Semiclassical} is devoted to a check of the S-matrix by taking the
semi-classical limit. Section~\ref{CP2} focusses on the symmetric space $\CP^2$ which
is somewhat different and simpler because the symmetry group is abelian. In this
case we are also able to test our conjectured S-matrix by using the
thermodynamic Bethe Ansatz. Finally in Section~\ref{Comparison} we
draw some conclusions.

\section{The Symmetric Space Sine-Gordon Theories}
\label{PohlmeyerR}

The starting point is a sigma model with target
space a symmetric space $F/G$. The group in the numerator $F$ admits 
an involution $\sigma$ whose stabilizer is the
subgroup $G$. Acting on the Lie algebra of $F$, the involution gives rise to the canonical decomposition
\EQ{
{\mathfrak f} = {\mathfrak g} \oplus {\mathfrak p}
\quad \text{with} \quad 
[{\mathfrak g},{\mathfrak g}]\subset {\mathfrak g}\>, 
\quad [{\mathfrak g},{\mathfrak p}]\subset 
{\mathfrak p}\>, 
\quad [{\mathfrak p},{\mathfrak p}]\subset {\mathfrak g}\>,
\label{CanonicalDec}
}
where ${\mathfrak g}$ and ${\mathfrak p}$ are the $+1$ and $-1$ eigenspaces of $\sigma$, respectively.
This allows us to formulate the symmetric space in terms
of a group element $\newf\in F$ constrained via
\EQ{
\sigma(\newf)= \newf^{-1}\>.
\label{Cond}
}
For the case of $\CP^{n+1}=SU(n+2)/U(n+1)$ 
we can describe the target space in terms
of $n+2$ complex homogeneous coordinates $\BZ$ with the identification 
$\BZ\sim\lambda\BZ$, $\lambda\in\C^\star$.
The map from the space $\CP^{n+1}$ to the group field realized in the
$n+2$-dimensional defining representation of $SU(n+2)$ is given by
\EQ{
\newf = \theta \left( I - 2\> \frac{\BZ \BZ^\dagger}{|\BZ|^2} \right)\>,
}
where $\theta= \text{diag}(-1,1,\ldots,1)$ implements the involution
\EQ{
\sigma(\newf)= \theta \newf \theta\>.
}
The subgroup $G=U(n+1)$ then consists of elements $\newf$ of the form
\EQ{
\newf=\left(\begin{array}{cc}e^{i\phi}&\B0\\
    \B0&W\end{array}\right)\in SU(n+2)\>,
}
with $W\in U(n+1)$ and $e^{-i\phi}=\det W$.

The Lagrangian of the sigma model is 
\EQ{
\LAG= -\Tr \big( {\cal J}_\mu {\cal J}^\mu\big)\quad \text{with}
\quad {\cal J}_\mu=\partial_\mu \newf \newf^{-1}\>,
\label{Lag}
}
whose equations-of-motion are
\EQ{
\partial_\mu {\cal J}^\mu=0\>.
\label{p1}
}
The conserved current
$ {\cal J}_\mu$ corresponds to the
global symmetry transformation\footnote{The Lagrangian density~\eqref{Lag} is invariant under the global transformations $\newf\to U \newf V$ for any $U,V \in F$. However, this symmetry is reduced by the constraint~\eqref{Cond} so that the symmetric space sigma model is invariant only under~\eqref{Global}.
}
\EQ{
\newf\to U \newf \sigma(U^{-1})\>,\qquad
U\in F\>,
\label{Global}
}
with a conserved Noether charge
\EQ{
{\cal Q} = \int_{-\infty}^{+\infty} \partial_0\newf \newf^{-1}
}
that takes values in the Lie algebra of $F$.

In the case of the $\CP^{n+1}$
sigma model, the Pohlmeyer reduction involves imposing the
conditions~\cite{Hollowood:2009tw,Miramontes:2008wt}\footnote{In our notation, $x_\pm = t\pm x$ and $\partial_\pm
  =\frac{1}{2}(\partial_t\pm\partial_x)$.} 
\EQ{
\partial_\pm \newf \newf^{-1} =\mu f_\pm \Lambda f_\pm^{-1}\>,
\label{PohlConst}
}
where $f_\pm\in F$ and 
\EQ{
\Lambda=\left(\begin{array}{cc|c}0&-1&\B0\\1&0&\B0\\\hline \B0&\B0&\B0\end{array}\right)\>.
}
Here, $\mu$ is an arbitrary mass scale and in most of the following we
shall set $\mu=1$ with the understanding that it can be re-introduced in
order to reconcile the dimensions.
$\Lambda$ is the unique element, up to conjugation, of the the $-1$
eigenspace $\mathfrak p$ of the Lie algebra
of $SU(n+2)$.

Notice that the Pohlmeyer
constraints break the Lorentz and conformal invariance of the sigma
model. The key observation is that the Lorentz invariance can be recovered
by a re-formulation of the constrained system as
a relativistically
invariant, massive and integrable theory: this is the 
symmetric space sine-Gordon
theory~\cite{Pohlmeyer:1975nb,Bakas:1995bm,Miramontes:2008wt}. The degree-of-freedom of the SSSG
theory is the $G$-valued field
\EQ{
\gamma=f_-^{-1}f_+
\label{xxd1}
}
which satisfies the SSSG equations
\EQ{
\big[\partial_++\gamma^{-1}\partial_+\gamma+\gamma^{-1}A_+^{(L)}\gamma-\tfrac{1}{2}\Lambda\>,\;
\partial_-+A_-^{(R)}-\tfrac{1}{2}\gamma^{-1}\Lambda\gamma\big]=0\>.
\label{www1}
}
Here, the quantities $A^{(L)}_+$ and 
$A^{(R)}_-$ can be interpreted as lightcone components of gauge fields
associated to a $H_L\times H_R$ gauge symmetry under which
\EQ{
f_\pm\longrightarrow f_\pm h_\pm^{-1}\ ,
}
where $h_\pm$ are local group elements in the subgroup
$H\subset G$, the subgroup of $G\subset F$ that commutes with $\Lambda$. In the present case $H=U(n)$ and its elements are of the form
\EQ{
\left(\begin{array}{c|c}e^{i\phi} \Bone&\B0\\\hline \B0& M\end{array}\right)\in SU(n+2)
}
with $M\in U(n)$ and $e^{-2i\phi}=\det M$.
Under this symmetry
\EQ{
\gamma\longrightarrow h_- \gamma h_+^{-1}\ 
\label{gaugeLR1}
}
and 
\EQ{
A_-^{(R)}\longrightarrow h_+\big(A_-^{(R)}+\partial_-\big)h_+^{-1}\ ,\qquad
A_+^{(L)}\longrightarrow h_-\big(A_+^{(L)}+\partial_+\big)h_-^{-1}\ .
\label{gaugeLR2}
}

For general non-abelian $H$, a 
Lagrangian formalism can be found by identifying ${\cal
  A}_-=A_-^{(R)}$ and ${\cal A}_+=A_+^{(L)}$ as the two lightcone
components of a gauge field, and by imposing the constraints~\cite{Bakas:1995bm}
\SP{
&\Big[\gamma^{-1}\partial_+\gamma
+\gamma^{-1}{\cal A}_+\gamma\Big]_{\mathfrak h}={\cal A}_+\ ,\\
&\Big[-\partial_-\gamma\gamma^{-1}
+\gamma{\cal A}_-\gamma^{-1}\Big]_{\mathfrak h}={\cal A}_-\>,
\label{gco2}
}
where the projection is onto the Lie algebra of $H$. These conditions
can be viewed as a set of partial gauge fixing
conditions~\cite{Grigoriev:2007bu,Miramontes:2008wt}.
They reduce the $H_L\times H_R$ gauge symmetry~\eqref{gaugeLR1} to the
$H$ vector subgroup\footnote{Note that it is also possible to gauge
  the axial vector subgroup of the overall $U(1)$ subgroup of
  $H$, whilst still gauging the vector subgroup of the non-abelian
  factor $SU(n)$. This gives rise to a different Lagrangian
  formulation of the theory for which the solitons carry a $U(1)$
  Noether charge \cite{Hollowood:2009tw}. The two formulations are
  related by a kind of T-duality \cite{Miramontes:2004dr}.} 
\EQ{
\gamma\longrightarrow U\gamma U^{-1}\>, \qquad U\in H\>,
\label{gaugeH}
}
under which ${\cal A}_\mu$ transforms as a gauge connection:
\EQ{
{\cal A}_\mu\longrightarrow U\big({\cal
  A}_\mu+\partial_\mu\big)U^{-1}\ .
\label{gaugeHA}
}
The gauge-fixed equations-of-motion (with $\mu$ re-introduced) are then
\EQ{
\big[\partial_++\gamma^{-1}\partial_+\gamma
+\gamma^{-1}{\cal A}_+\gamma,\partial_-+{\cal A}_-\big]
=\frac{\mu^2}{4}[\Lambda,
\gamma^{-1}\Lambda\gamma]
\label{eom3}
}
and these follow as the 
equations-of-motion of the Lagrangian density
\SP{
\LAG&=\LAG_{WZW}(\gamma)+\frac1{2\pi}\Tr\Big(-{\cal 
A}_+\partial_-\gamma
\gamma^{-1}+{\cal A}_-\gamma^{-1}\partial_+\gamma\\
&+\gamma^{-1}{\cal A}_+\gamma{\cal A}_- -{\cal A}_+{\cal A}_-
-\frac{\mu^2}{4}\Lambda
\gamma^{-1}\Lambda\gamma\Big)\ ,
\label{ala}
}
where $\LAG_{WZW}(\gamma)$ is the usual WZW Lagrangian density 
for $\gamma$. In fact this theory is the
gauged WZW model for $G/H$ deformed by 
the last term which is a potential. 
Notice that the partial gauge-fixing constraints \eqref{gco2} now appear as
the equations-of-motion of the gauge connection. 
The coupling of the
theory is the level of the WZW part of the action which we denote by
the integer $k$.

At the classical level, the deformed WZW model has a degenerate vacuum
which one can identify with constant elements $\gamma_v\in H$ modulo gauge
transformations: $\gamma_v\sim U\gamma_v U^{-1}$, $U\in H$. In other words, there is a classical vacuum moduli space that is the Cartan torus
of $H=U(n)$. However, the putative massless fluctuations in field
directions in the moduli space turn out to have singular kinetic
terms. 

As an example, we can consider the case of $\CP^2$ discussed in
detail in \cite{Hollowood:2009tw}.
In this case
$H=U(1)$ is abelian and we can use both vector or axial gauging to achieve a
Lagrangian formulation. For present purposes we discuss the vector
gauged model which generalizes to the non-abelian cases.
We can gauge fix the vector symmetry by choosing a gauge slice of the form
\EQ{
\gamma=\MAT{e^{i\psi/2}&0&0\\ 0&\cos\theta
  e^{i(\varphi+\psi/2)}&\sin\theta e^{-i\psi/2}\\
0&-\sin\theta e^{i\psi/2}&\cos\theta e^{-i(\varphi+\psi)}}\ .
\label{ksa}
}
We then solve 
the conditions \eqref{gco2} for
${\cal A}_\mu$ and then insert these into the Lagrangian to give an
effective Lagrangian for the physical degrees-of-freedom
\EQ{
\LAG=\partial_\mu\theta\partial^\mu\theta+\frac14\partial_\mu\psi
\partial^\mu\psi+\cot^2\theta\partial_\mu(\psi+\varphi)\partial^\mu(
\psi+\varphi)+2  \cos\theta\cos\varphi\ .
}
Notice that the vacuum is degenerate with 
$\theta=\varphi=0$ and $0\leq\psi<4\pi$, but note that the kinetic
term for $\psi$ is singular due to the $\cot\theta$ pre-factor. At
this stage we can only conclude that a conventional approach to
quantization via perturbation theory is likely to be unconventional
\cite{Grigoriev:2007bu,Hoare:2009rq,Hoare:2009fs}.

\section{The Classical Magnons/Solitons}
\label{Dressing}

The non-relativistic system consisting of the 
original $F/G$ sigma model subject to the Pohlmeyer constraints
has lump-like solutions known as giant magnons. These solutions have a
relativistic soliton ``avatar'' that satisfies the SSSG equations: these
are solitons in the form of kinks carrying topological charges 
of the deformed WZW theory. 

The map between the magnons and solitons is complicated. However,
in  \cite{Hollowood:2009tw} it was shown 
how the dressing method, applied to magnons in
\cite{Spradlin:2006wk}, can be used to construct both
the magnons and their soliton avatars simultaneously without the
need to map one into the other. Here, we
briefly review the construction for the $\CP^{n+1}$ in order to
describe the solitons and, importantly, to reveal their internal
structure in some detail. 

The dressing transformation method makes use of the associated linear system
\EQ{
\partial_\pm \Psi(x;\lambda) =\frac{\partial_\pm \newf \newf^{-1}}{1\pm\lambda}\Psi(x;\lambda)\>, \qquad
\Psi(x;\infty)=I
\>, \qquad \newf(x)= \Psi(x;0)\>,
\label{LinearP}
}
whose integrability conditions are equivalent to the equations of motion of the sigma model. For $\CP^{n+1}$, the solutions $\Psi(x;\lambda)$ have to satisfy the two conditions
\EQ{
\Psi^{-1}(x;\lambda) = \Psi^\dagger(x;\lambda^\ast) \>,\qquad
\Psi(x;1/\lambda)=\newf\theta\Psi(x;\lambda)\theta \>,
\label{CondPsi}
}
which ensure that $\newf^{-1}=\newf^\dagger$ and that the constraint~\eqref{Cond} is satisfied. Then, the dressing transformation involves constructing a new solution $\Psi$ of the linear system of the form
\EQ{
\Psi(x;\lambda)= \chi(x;\lambda) \Psi_0(x;\lambda)
\label{DressingT}
}
in terms of an old one $\Psi_0(x;\lambda)$, 
which can be chosen to be the ``vacuum'' solution
\EQ{
\Psi_0(x;\lambda)=\exp\Big[ \Big(\frac{x_+}{1+\lambda} +\frac{x_-}{1-\lambda}\Big) \Lambda\Big]\>.
}
In terms of the homogeneous coordinates, it corresponds to $\BZ_0=
(\cos t, -\sin t, \B0)$. This vacuum solution, on the sigma model
side, represents the motion of a point-like string on the target space
$\CP^{n+1}$ at the speed of light. 

Following~\cite{Harnad:1983we}, the general form of the ``dressing factor'' is
\EQ{
\chi(\lambda)= 1+ \sum_i \frac{Q_i}{\lambda-\lambda_i}\>, \qquad
\chi^{-1}(\lambda)= 1+ \sum_i \frac{R_i}{\lambda-\mu_i}\>,
}
where the residues are rank-$1$ matrices of the form
\EQ{
Q_i= \BX_i \BF_i^\dagger\>, \qquad
R_i= \BH_i \BK_i^\dagger
}
for vectors $\BX_i$, $\BF_i$, $\BH_i$, and $\BK_i$. For $\CP^{n+1}$, they are given by
\SP{
&
\BX_i \Gamma_{ij}=\BH_j\>, \qquad
\BK_i \big(\Gamma^\dagger\big)_{ij} = -\BF_j\>, \qquad
\Gamma_{ij}=\frac{\BF_i^\dagger \BH_j}{\lambda_i-\mu_j}\>,\\[5pt]
&
\BF_i = \Psi_0(\lambda_i^\ast) \Bvarpi_i\>, \qquad
\BH_i = \Psi_0(\mu_i) \Bpi_i\>,
\label{wer}
}
where $\Bvarpi_i$ and $\Bpi_i$ are complex constant $n+2$~dimensional vectors. The allowed number of poles and their positions are constrained by the conditions~\eqref{CondPsi}. They imply that $\mu_i=\lambda_i^\ast$ and, moreover, that the poles $\{\lambda_i\}$ must come in pairs $(\lambda_i,\lambda_{i+1}=1/\lambda_i)$. In addition, $\Bpi_i=\Bvarpi_i$ and, for each pair,
\EQ{
\Bvarpi_{i+1}=\theta\Bvarpi_i\>.
}

One the main results of \cite{Hollowood:2009tw} is that the dressing
transformation not only produces the giant magnons but also directly 
the soliton ``avatars'' of the related SSSG in the form
\EQ{
\gamma={\cal F}_0^{-1/2}\chi(+1)^{-1}\chi(-1){\cal
  F}_0^{1/2}\ ,
\label{xxd}
}
with $A^{(L)}_+=A^{(R)}_-=0$. This expression automatically satisfies the
constraints \eqref{gco2}
\EQ{
\Big[\gamma^{-1}\partial_+\gamma\Big]_{{\mathfrak
    h}}=\Big[\partial_-\gamma\gamma^{-1}\Big]_{{\mathfrak h}}=0\ .
\label{gco}
}

The basic soliton for the $\CP^{n+1}$ case is obtained by taking a
solution with a single pair of poles $\{\xi,1/\xi\}$ where we parametrize
$\xi=re^{ip/2}$. The ``dressing data'' involves the complex $n+2$
vector $\Bvarpi$, with
$\Bvarpi_1=\Bvarpi$ and $\Bvarpi_2=\theta\Bvarpi$. The various choices
that can be made are discussed at length in~\cite{Hollowood:2009sc}. The basic magnon,
or its soliton avatar, is obtained by taking
\EQ{
\Bvarpi=(1,i,\BOmega)\ ,
}
where $\BOmega$ is a complex $n$-dimensional vector subject to
$|\BOmega|=1$. The magnons, or solitons, only depend on $\BOmega$ up to
a phase and so the lump has an internal collective coordinate valued in
$\CP^{n-1}$. 

The data $\{\xi=re^{ip/2},\BOmega\}$ (where we will implicitly identify
$\BOmega\sim e^{i\alpha}\BOmega$) determines the rapidity and the
charges of the magnon and its soliton avatar. The rapidity of the
magnon and of the soliton are of course equal---they are the same object
viewed from two different perspectives---and is determined by
\EQ{
\tanh\vartheta= \frac{2r}{1+r^2}\cos\frac p2\>.
\label{rap}
}
The $SU(n+2)$ Noether charge of the magnon is\footnote{The charge ${\cal Q}$ is defined relative to the vacuum
  $\Delta{\cal Q}={\cal Q}-{\cal Q}_0$.}
\EQ{
\Delta{\cal Q} = J_\Lambda \Lambda + J_H h_\BOmega\>,\quad J_\Lambda=-\frac{1+r^2}{r}\big|\sin\frac{p}{2}\big|\>, \quad
J_H= -\frac{1-r^2}{r}\big|\sin\frac{p}{2}\big|\>,
\label{NewCharge2}
}
where 
\EQ{
h_\BOmega=i\left(\begin{array}{c|c}\boldsymbol{1} & \B0
\\ \hline
\B0 &-2\BOmega\BOmega^\dagger 
\end{array}\right)
\label{defh}
}
is one of the infinitesimal generators of $H=U(n)$, which is the subgroup of elements of $G=U(n+1)$ that commute with $\Lambda$.
These charges satisfy the relation
\EQ{
-J_\Lambda = \sqrt{J_H^2 + 4\sin^2\frac{p}{2}}\>.
\label{Dispersion}
}
In the AdS/CFT context~\cite{Hofman:2006xt,DyonicGM}, $J_\Lambda$ and $J_H$ are identified, up to scaling, with $\Delta-\frac{1}{2}J$ and $Q$, respectively, where $\Delta$ is the scaling dimension of the associated operator in the CFT, and $J$ and $Q$ are two conserved $U(1)$ $R$-charges:\footnote{To be specific, we use the same normalization as~\cite{Abbott:2009um}.}
\EQ{
\Delta -\frac{1}{2}J = -\sqrt{\frac{\lambda}{2}}\> J_\Lambda\>, \qquad
\frac{1}{2}Q = \sqrt{\frac{\lambda}{2}}\> J_H\>,
}
where $\lambda$ is the 't~Hooft coupling. Then,~\eqref{Dispersion} becomes the celebrated (non-relativistic) dispersion relation
\EQ{
\Delta -\frac{1}{2}J = \sqrt{\frac{1}{4}Q^2 + 2\lambda\sin^2\frac{p}{2}}\>.
}

In the SSSG theory, the soliton is a relativistic kink with a topological charge 
\EQ{
\gamma(-\infty)^{-1}\gamma(\infty)
=\exp(-2q h_\BOmega)\ ,
}
where $h_\BOmega$ is the Lie algebra element \eqref{defh}
and\footnote{If we define $X^\pm=re^{\pm ip/2}$, which are convenient
  variables from the magnon side, and $Z^\pm=e^{\vartheta\pm iq}$,
  which are convenient variables from the soliton side, then
  $X^\pm=(Z^\pm-1)/(Z^\pm+1)$ and $Z^\pm=(1+X^\pm)/(1-X^\pm)$.} 
\EQ{
\tan q= \frac{2r}{1-r^2} \sin\frac{p}{2}\ .
}
The mass of the soliton is then
\EQ{
m=\frac{2k\mu}\pi\left|\sin q\right|\ .
\label{solm}
}
The inequivalent solutions are obtained by restricting $0\leq
p\leq 2\pi$. The charge $q$ can then be chosen to lie $-\frac\pi2
\leq q\leq\frac\pi2$ and solitons with $r<1$ have charge $0\leq q\leq\frac\pi2$ and those
with $r>1$ have charge $-\frac\pi2\leq q\leq0$. There is a
notion of charge conjugation that takes $r\to1/r$, or $\xi\to1/\xi^*$,
and $q\to-q$. Notice that  $q$ is only defined modulo $\pi$. Therefore, $q=+ \frac\pi2$ and  $q=- \frac\pi2$ actually correspond to the same solutions, which are those obtained with $r=1$. 

The internal collective coordinate
$\BOmega$ plays a different role for magnons and solitons. For the
former, there is an $H\subset F$ Noether symmetry under which $\BOmega$ transforms
as a vector. One can think of $\BOmega$ as a kind of angular momentum
and in the quantum theory one finds that states will come in
representations of $H$. For the solitons, $\BOmega$ encodes the topological kink
charge rather than a Noether charge. 
Classically the perturbed gauged WZW model has a vacuum which we can
identify with a constant element $\gamma_v\in H$ modulo gauge
transformations: $\gamma_v\sim U\gamma_v U^{-1}$, $U\in H$; 
in other words, there is a vacuum moduli space that is the Cartan torus
of $H=U(n)$. Consequently, the topological charge should be thought of
as taking values in a Cartan subalgebra of $H$. In the quantum theory,
we will find that the classical moduli space 
$\CP^{n-1}$ becomes a ``fuzzy'' space with
non-commuting coordinates and once again the quantum states form
representations of $H$, or more precisely its $q$ deformation the
quantum group $U_q(H)$, where $q$ is determined by $k$.

For multi-soliton solutions it is not generally possible
to use gauge transformations to take the topological charge of each
soliton into the same Cartan subalgebra. From this perspective, the
dependence on $\BOmega$ simply corresponds to the freedom to choose
the Cartan subalgebra, and the scattering amplitudes will depend on
$\BOmega$. The situation simplifies for the special choices
$\BOmega^{(i)}_a =\delta_{ia}$, for $i=1,\ldots,n$, or in vector
language $\BOmega^{(i)}=\Be_i$, such that the corresponding  
solitons carry topological charges laying on the same Cartan
subalgebra. These special solutions play an important role later in
section~\ref{SemiC}
because it is particularly simple to relate their classical scattering
to the semi-classical limit of the quantum S-matrix.

\section{The Soliton S-matrix Conjecture}
\label{Smatrix}

Finding the S-matrix of an integrable field theory is never a direct
process: one has to use a variety of evidence in order to pin it down.
In an ideal world one would like to 
quantize the perturbed gauge WZW model from first
principles but this is not something that can be done with present understanding. 
Fortunately, there are plenty of clues and many other examples to
guide us. Firstly, integrable deformations
of WZW theories typically lead to S-matrices describing a system of
kinks. For example, there are integrable deformations of WZW models
associated to the symmetric spaces $G/H$ \cite{Fendley:1999gb} (unlike
the present situation where $G/H$ is {\it not\/} a symmetric
space). The deformation is these cases is
provided by the operator $\sum_a J_a\bar J_a$, with a sum over the
components of the currents in $G$ but not in $H$. Then,
the spectrum consists of a set of kinks which interpolate a finite set of vacua
associated to the irreducible representations of the symmetry
group $G$ of
level $\leq k$, where $k$ is the level of the WZW model. In this case
the kinks have topological charges which are weights of the
fundamental (anti-symmetric) representations of $G$, and the S-matrix elements involve
the trigonometric solution of the Yang-Baxter equation associated to
the quantum group $U_q(G)$ with a deformation parameter
$q=-\exp(i\pi/(k+h))$, where $h$ is the dual Coxeter number of $G$.
The fact that the S-matrix involves kinks seems to be related at a
fundamental level to a basis of 
quasi-particles known as spinons in the original coset CFT
\cite{Bouwknegt:1999ip,Guruswamy:1999cb,Bouwknegt:1994bk}. 

We will also find that the S-matrix of the symmetric space sine-Gordon
theories described as deformations of the $G/H$ WZW model are kinks
which interpolate a set of vacua which are associated to the
representations of $H=U(n)$ of level $\leq k$. Notice that in this
case the symmetry group is $H$ rather than $G$ since the latter
symmetry is
broken by the deforming potential.
The S-matrix will also involve the trigonometric solution of the Yang-Baxter equation
associated to $U_q(H)$. The main difference is that
the spectrum will involve the symmetric representations rather
than the anti-symmetric ones. It seems natural that these
theories should have a quantum group symmetry much like the
sine-Gordon theory itself whose kinks have an $U_q(SU(2))$ symmetry
\cite{Ahn:1990gn}. For generic $q$ the representations of $U_q(H)$ are simply deformations
of those of $H$, however, in
the present case $q$ is a root of unity and this means that the set of
representations is restricted in a way that is crucial to the
construction of the S-matrix. S-matrices associated to trigonometric
solutions of the Yang-Baxter equation have been considered in the
past \cite{Ahn:1990gn,deVega:1990av,Hollowood:1990qn,Hollowood:1992sy,Gepner:1993eg,Hollowood:1993fj}
the main difference with the present case is that those S-matrices
involved the anti-symmetric representations.

For $q$ a root of unity, it is most appropriate to use the
restricted-solid-on-solid (RSOS) picture for which the states of the
theory are kinks. The kinks interpolate between a
discrete set of vacuum states which are identified with the
irreducible representations of $SU(n)$ of level $\leq k$,
which we denote $\Lambda^*(k)$.
Concretely these are associated
to Young Tableaux whose width is restricted to be $\leq k$, or the
set of vectors $\sum_{i=1}^{n-1}a_i\Be_i$, with 
\EQ{
k\geq a_1\geq a_2\geq\cdots\geq a_{n-1}\geq0\ ,
\label{wres}
}
where the $\Be_i$'s provide the set of weights of the vector representation of $SU(n)$ (see~\ref{sigma1}). A kink with rapidity $\vartheta$ is then denoted $K_{\Ba\Bb}(\vartheta)$
for $\Ba,\Bb\in\Lambda^*(k)$. The topological charges of a kink
$\Ba-\Bb$ are weights associated to the Cartan elements of
$U_q(SU(n))$.\footnote{The overall $U(1)$ subgroup of $H$ is trivially
  represented on the kinks.} Note
that these elements will commute with the S-matrix and the topological
charge is conserved. Notice also that the set of vacua describe a kind of
discretization of the Cartan torus of $H$, the classical vacuum moduli
space, that is recovered in the limit $k\to\infty$.

In an integrable field theory the
complete S-matrix is then determined by the S-matrix for the $2\to2$ processes
\EQ{
K_{\Ba\Bc}(\vartheta_1)+K_{\Bc\Bd}(\vartheta_2)\rightarrow K_{\Ba\Bb}
(\vartheta_2)+K_{\Bb\Bd}(\vartheta_1)\ .
\label{KP}
}
We will find that the topological charge $\Ba-\Bb$ of a kink
$K_{\Ba\Bb}(\vartheta)$ have to be weights of one of the symmetric
representations of $SU(n)$ with Young Tableau $[a]$, or their
conjugates $[a,\ldots,a]=[a^{n-1}]$.\footnote{We 
use the label $[a_1,a_2,\ldots,a_{n-1}]$
  for the representation of $SU(n)$ with highest weight
  $\sum_ia_i\Be_i$. 
In a Young Tableaux $a_i$ gives the
  number of boxes in $i^\text{th}$ row and $a_1\geq a_2\geq\cdots\geq
  a_{n-1}\geq0$. The quantum group further restricts $k\geq a_1$.} 
With $q$ a root of unity, $q=-\exp(i\pi/(n+k))$, the quantum group restriction 
means that $a=1,\ldots,k$ only.
We denote the set of weights in the representation $[a]$ and
$[a^{n-1}]$ as $\Sigma_{[a]}$ and $\Sigma_{[a^{n-1}]}$, respectively.
We will identify the overall $U(1)$ kink charge (not to be confused
with the quantum group deformation parameter) as equal to $q=\pm\pi
a/N$, for kinks and anti-kinks, respectively, for an
integer $N$ that will be identified as we proceed. The mass of 
a kink with a topological charge in $\Sigma_{[a]}$ or
$\Sigma_{[a^{n-1}]}$ follows from the classical formula \eqref{solm}
\EQ{
m_a=M\sin\left({\pi a\over N}\right),\quad a=1,2,\ldots,k\ ,
\label{AMAS}
}
where $M$ is an overall renormalized mass scale.

The S-matrix elements are constructed from the 
trigonometric solutions of the Yang-Baxter equation associated to a 
certain deformation of
the universal enveloping algebra of the
Lie algebra known as a quantum group \cite{JIM,DRIN}: in the
present context $U_q(SU(n))$.
The solutions can be thought of as intertwiners between
tensor products of representations of the algebra:
\EQ{
\rc(\vartheta):\ U(\vartheta_1)\otimes V(\vartheta_2)
\rightarrow V(\vartheta_2) \otimes U(\vartheta_1)\ ,
\label{INT}
}
where $\vartheta=\vartheta_1-\vartheta_2$ 
is the (additive) spectral parameter which we will
later identify with the rapidity. Such an $R$-matrix
has a spectral decomposition \cite{JIM}
\EQ{
\rc(\vartheta)=\sum_{W\subset U\otimes V}
\rho_\lambda(\vartheta){\Bbb P}_W,
}
where ${\Bbb P}_W$ is a quantum group invariant homomorphism from
$U\otimes V$ to $V\otimes U$ with the
property that $\sigma{\Bbb P}_W$ is a projection onto $W\subset
U\otimes V$,
where $\sigma: v\otimes u\mapsto u\otimes
v$, for $u\in U$ and $v\in V$,  is the permutation.
It is important that, in the context of the quantum group, the tensor product is a subset of the tensor product of the group.
In the following we shall switch between the language of spectral
decompositions and the RSOS picture where necessary.

{\sl The basic S-matrix elements}

To start with we consider
the solutions associated to the vector representation of the algebra
$\rc_{11}(\vartheta)$. The set of weights of the vector representation are
\EQ{
\Sigma_{[1]}=\big\{\Be_1,\ldots,\Be_n\big\}\ ,
\label{sigma1}
}
where the $\Be_i$'s are a set of vectors with $\Be_i\cdot
\Be_j=\delta_{ij}-1/n$ in an $n-1$-dimensional space, and $\sum_{i=1}^n\Be_i=0$.

The solution of the YBE is labelled by four weights of the algebra:
\EQ{
\rc_{11}\left(\left.\ARR{\Ba&\Bb\\ \Bc&\Bd}\right\vert\vartheta\right),
\quad \Ba,\Bb,\Bc,\Bd\in\Lambda^\star(k),
}
with the property that $\rc_{11}$ is only non-zero if $\Bc-\Ba$, $\Bd-\Bc$, 
$\Bb-\Ba$ and $\Bd-\Bb$
are in $\Sigma_{[1]}$. For completeness we now write down the explicit 
solutions following \cite{JMO} (see
also the review \cite{WDA}). In the following $\omega$ is a constant which
is related to the deformation parameter of the quantum group
$q=-e^{i\omega}$ and so $\omega=\frac\pi{n+k}$.
For convenience we introduce for $\Ba\in\Lambda^\star(k)$ and
$\Bmu,\Bnu\in\Sigma_{[1]}$
\EQ{
a_\Bmu=\omega(\Ba+\Brho)\cdot\Bmu\ ,~~~~~a_{\Bmu\Bnu}=a_\Bmu-a_\Bnu\ ,
}
where $\Brho$ is the sum of the fundamental weights of the algebra.\footnote{These
are the vectors $\Bomega_i=\Be_1+\cdots+\Be_i$, for $i=1,\ldots,n-1$.} 

With a suitable choice of overall normalization, the solution is
\EQ{
&\rc_{11}\left(\left.\ARR{\Ba&\Ba+\Be_i\\ \Ba+\Be_i&\Ba+2\Be_i}\right\vert
  \vartheta\right)=1\ ,\\
&\rc_{11}\left(\left.\ARR{\Ba&\Ba+\Be_i\\ \Ba+\Be_i&\Ba+\Be_i+\Be_j}
\right\vert\vartheta\right)=
\frac{\sin(a_{\Be_i\Be_j}+i\lambda\vartheta)\sin\omega}{\sin(a_{\Be_i\Be_j})
\sin(\omega-i\lambda\vartheta)}\ , \\
&\rc_{11}\left(\left.\ARR{\Ba&\Ba+\Be_j\\ \Ba+\Be_i&\Ba+\Be_i+\Be_j}\right\vert\vartheta\right)=
{\sin(i\lambda\vartheta)\over\sin(\omega-i\lambda\vartheta)}
\left({\sin(a_{\Be_i\Be_j}+\omega)\sin(a_{\Be_i\Be_j}-\omega)\over
\sin^2(a_{\Be_i\Be_j})}\right)^{1/2}\ .
\label{hj2}
}
The solution satisfies the unitarity condition
\EQ{
\sum_{\Bz}\rc_{11}\left(\left.\ARR{\Ba&\Bz\\ \Bc&\Bd}
\right\vert\vartheta\right)
\rc_{11}\left(\left.\ARR{\Ba&\Bb\\ \Bz&\Bd}\right\vert -\vartheta\right)=
\delta_{\Bb\Bc}\ .
\label{CR}
}

The solution of the YBE equation written above 
naturally leads to an $S$-matrix for the two $[1]$ kink process
once multiplied by a suitable scalar factor,
\EQ{
S_{11}\left(\left.\ARR{\Ba&\Bb\\ \Bc&\Bd\\}\right\vert
  \vartheta\right)=
X_{11}(\vartheta)Y_{11}(\vartheta)
\rc_{11}\left(\left.\ARR{\Ba&\Bb\\ \Bc&\Bd\\}\right\vert
  \vartheta\right)\ .
\label{SMAT}
}
The fact that we split the scalar factor into 2 pieces
$X_{11}(\vartheta)$ and $Y_{11}(\vartheta)$ is for
convenience: the first factor will contain all the bound-state poles
on the physical strip while the second is needed to satisfy untarity
and crossing. 
Both factors must be chosen so that the $S$-matrix axioms are
satisfied and the semi-classical limit of the S-matrix is
consistent. For instance, assuming that $X_{ab}(\vartheta)$ satisfies
unitarity and crossing separately, we see from \eqref{CR} that
unitarity requires
\EQ{
Y_{11}(\vartheta)Y_{11}(-\vartheta)=1\ .
\label{nn1}
}

{\sl The bootstrap}

We now describe how to build up the full S-matrix from this basic one
by applying the bootstrap. The idea is that simple poles on
the physical sheet in rapidity space, $0\leq\IM\vartheta\leq\pi$, are interpreted as the
exchange of a bound-state in either the direct or crossed channel.
If we look at the spectral decomposition of the basic $R$-matrix then
\EQ{
\rc_{11}(\vartheta)={\Bbb P}_{[2]}-{\sinh(\lambda\vartheta-i\omega)
\over\sinh(\lambda\vartheta+i\omega)}{\Bbb P}_{[1^2]},
\label{APR}
}
where ${\Bbb P}_{[2]}$ and ${\Bbb P}_{[1^2]}$ are the quantum group
invariant projectors which appear in the tensor product of two vector
representations. The idea of the bootstrap is that kinks with
topological charges in $\Sigma_{[2]}$ 
will appear as a bound-state provided $X_{11}(\vartheta)$ has a simple
pole at a rapidity difference that is fixed by the formula
\EQ{
m_2^2=m_1^2+m_1^2+2m_1^2\cos\vartheta\ ,
}
giving $\vartheta=2i\pi/N=i\omega/\lambda$. This fixes 
\EQ{
\lambda=\frac{N}{2(k+n)}\ ,
}
so that the residue of the pole is proportional to
${\mathbb P}_{[2]}$. This condition on
$X_{11}(\vartheta)$ is not enough to complete fix it. This kind of
situation is common in an integrable field theory, S-matrix can often
only be determined up to ``CDD factors'', that is functions which are
analytic on the physical strip. In the present case, we will simply
postulate a form for $X_{11}(\vartheta)$ which is consistent with the
semi-classical limit that we discuss later:\footnote{In principle, the
  semi-classical limit and bootstrap 
allows for a more general expression where 
the arguments of the hyperbolic functions are scaled by some
function which $\to1$ as $k\to\infty$. For example the scaling could
be by $\lambda^{-1}$. However, such a choice would differ only by
CDD factors from the one we chose. 
Such ambiguities would be pinned down by a TBA calculation of
the central charge.}
\EQ{
X_{11}(\vartheta)=
\frac{\sinh(\frac{\lambda\vartheta}2+\frac{i\omega}2)}
{\sinh(\frac{\lambda\vartheta}2-\frac{i\omega}2)}
\frac{\cosh(\frac{\lambda\vartheta}2+\frac{i\omega}4)}
{\cosh(\frac{\lambda\vartheta}2-\frac{i\omega}4)}\ .
}
The first quotient here is strictly-speaking the minimum that is
necessary since it provides the simple pole. The second factor is a
CDD factor that we will later find is necessary to produce the correct
semi-classical limit. We cannot rule out further CDD factors that have
a trivial semi-classical limit.

The S-matrix
elements for the bound-state kinks with the fundamental kinks then
follows from the bootstrap equations
\EQ{
S_{21}(\vartheta)=S_{11}(\vartheta+\tfrac{i\pi}N)S_{11}(\vartheta-\tfrac{i\pi}N)\ .
}
where the right-hand side is implicitly restricted to $[2]\times[1]$ in the tensor
product $[1]\times[1]\times[1]$.

The bootstrap then proceeds in a similar fashion to generate all the particles with
$a=1,\ldots,k$ transforming in representations $[a]$ with S-matrix
elements
\EQ{
S_{ab}(\vartheta)=X_{ab}(\vartheta)Y_{ab}(\vartheta)\rc_{ab}(\vartheta)\ ,
\label{hui}
}
where $\rc_{ab}(\vartheta)$ is the RSOS solution of the Yang-Baxter
equation for the product of representations $[a]\times[b]$. 
The tensor product $[a]\times[b]$ in the quantum group is only a
subset of the tensor product in the group itself:
\EQ{
[a]\times[b]=\bigoplus_{j=\text{max}(0,a+b-k)}^{\text{min}(a,b)}[a+b-j,j]\ .
\label{zx1}
}
The lower limit in here involves the
level $k$ and is a consequence of the quantum group structure at $q$ a
root of unity. The masses of the kinks determine that
$S_{ab}(\vartheta)$ should have a bound state pole at
$\vartheta=i\pi(a+b)/N$ corresponding to kinks with topological charge
in $[a+b]$. We must now verify that $S_{ab}(\vartheta)$ has this pole
and also that the residue is proportional to ${\mathbb P}_{[a+b]}$.

The bootstrap equations in general takes the form
\EQ{
S_{a+b,c}(\vartheta)=S_{ac}(\vartheta+\tfrac{i\pi
  b}N)S_{bc}(\vartheta-\tfrac{i\pi a}N)\ .
}
where the right-hand side is implicitly restricted to
$[a+b]\times[c]$ in the tensor product $[a]\times[b]\times[c]$. 
Applying the bootstrap equation recursively to the scalar factor gives 
\EQ{
X_{ab}(\vartheta)&=\prod_{j=0}^{a-1}\prod_{l=0}^{b-1}
X_{11}\Big(\vartheta+\frac{i\pi(a+b-2j-2l-2)}N\Big)\\
&=\prod_{j=1}^{\text{min}(a,b)}
\frac{(a+b-2j+2)_\vartheta(a+b-2j) _\vartheta[a+b-2j+1]_\vartheta}
{(2j-a-b-2) _\vartheta (2j-a-b) _\vartheta[2j-a-b-1]_\vartheta}\ ,
\label{uy3}
}
where we have defined for later use
\EQ{
(x) _\vartheta=\sinh\Big(\frac{\lambda\vartheta}2+\frac{i\omega x}4\Big)\ ,\qquad
[x]_\vartheta=\cosh\Big(\frac{\lambda\vartheta}2+\frac{i\omega x}4\Big)\ .
\label{dem}
}
Notice that $X_{ab}(\vartheta)$ does have a simple pole
at $\vartheta=i\pi(a+b)/N$ as required. It also has a simple pole at 
$i\pi|a-b|/N$ whose significance will emerge, and also
double poles at $i\pi(a+b-2j)/N$, $j=1,\ldots,\text{min}(a,b)-1$.

Now we turn to the $R$-matrix for 
$[a]\times[b]$ which has a spectral decomposition to match
\eqref{zx1}\footnote{These decompositions follow from the general
  technology involving the tensor product graph described in
  \cite{Zhang:1989sz}: see also Appendix A of \cite{Hollowood:1993fj}.}
\EQ{
\rc_{ab}(\vartheta)=
\sum_{j=\text{max}(0,a+b-k)}^{\text{min}(a,b)}(-1)^j\rho_{ab}^j(\vartheta){\Bbb
  P}_{[a+b-j,j]}
\label{nxx}
}
with
\EQ{
\rho_{ab}^j(\vartheta)=
\prod_{l=0}^{j-1}\frac{\sinh(\lambda\vartheta-i\omega(a+b-2l)/2)}
{\sinh(\lambda\vartheta+i\omega(a+b-2l)/2)}\ .
\label{qwq}
}
Although the quantum group structure fixes the spectral decomposition
of the $R$-matrix, it does not determine the overall
normalization which we have in hindsight fixed by setting
$\rho_{ab}^0(\vartheta)=1$. This must be fixed by solving the
bootstrap equation. 
Fortunately in the present context, since we are dealing with symmetric
representations, the normalization is
easy to fix by the following simple argument.
The two kinks with topological charge
$a\Be_i$ and $b\Be_i$ can only couple through the projector ${\Bbb
  P}_{[a+b]}$ because $(a+b)\Be_i$ can only be in $\Sigma_{[a+b]}$. 
Moreover, the basic $\rc_{11}$-matrix factor for the constituents
$\Be_i$ and $\Be_j$ is from \eqref{APR} unity and so applying the
bootstrap equation to these special states only we see that 
the $R$ matrix element for $a\Be_i$ with $b\Be_i$ must also be unity
and so, as we claimed,  $\rho_{ab}^0(\vartheta)=1$. 

Now that we have fixed the $R$-matrix, we can easily verify that at
the simple pole $\vartheta=i\pi(a+b)/N$ we have
\EQ{
\rc_{ab}(\vartheta)=\begin{cases}{\Bbb P}_{[a+b]} &a+b\leq k\\ 0 &a+b> k\end{cases}\ ,
\label{pin}
}
due to the factor $\sinh(\lambda\vartheta-i\omega(a+b)/2)$ in the
numerator of \eqref{qwq}.

From \eqref{uy3}, we see that there is another simple pole at
 $\vartheta=i\pi|a-b|/N$ which must be properly accounted for.
This will be identified with a bound
 state in the crossed channel. A consistent $S$-matrix must satisfy
crossing symmetry which requires that each kink $[a]$ has a charge
conjugate anti-kink with minus the topological charge, 
of the same mass, and transforming in the conjugate
representation with Young Tableau $[a^{n-1}]$. The crossing symmetry
relation then
gives the S-matrix elements for anti-kink/kink scattering as
\EQ{
S_{\bar ba}
\left(\left.\ARR{\Ba&\Bb\\ \Bc&\Bd\\}\right\vert \vartheta\right)
=S_{ab}\left(\left.\ARR{\Bc&\Ba\\ \Bd&\Bb\\}\right\vert i\pi- \vartheta\right)
,
}
Notice here that on the right-hand side
the topological charges $\Bb-\Bd$ and $\Ba-\Bc\in\Sigma_{[b]}$ 
whereas on the left-hand side
 $\Bd-\Bb$ and $\Bc-\Ba\in\Sigma_{[b^{n-1}]}$.
The cross-channel pole in $S_{ab}(\vartheta)$ at 
$\vartheta=i\pi|a-b|/N$ is then interpreted as a direct channel
pole at $\vartheta=i\pi-i\pi|a-b|/N$ for $[b^{n-1}]\otimes[a]$
scattering. If $a>b$ these must be
kinks in representation $[a-b]\subset[b^{n-1}]\times[a]$ 
which appears in the tensor product, while if $a<b$ they are
anti-kinks $[(b-a)^{n-1}]\subset[b^{n-1}]\times[a]$. 
For overall consistency, we require
that the $R$-matrix for $[b^{n-1}]\otimes[a]$, which we denote
$\rc_{\bar ba}(\vartheta)$, must be proportional to the projector
${\Bbb P}_{[a-b]}$, if $a>b$, and ${\Bbb P}_{[(b-a)^{n-1}]}$, if
$a<b$. The spectral decompositions are, firstly for
$a>b$,
\EQ{
R_{\bar ba}(\vartheta)=\Phi_{\bar ba}(\vartheta)
\sum_{j=0}^{b}(-1)^j\rho_{\bar ba}^j(\vartheta)
{\Bbb P}_{[a-b+2j,j^{n-2}]}\ ,
}
with 
\EQ{
\rho_{\bar ba}^j(\vartheta)=
\prod_{l=0}^{j-1}\frac{\sinh(\lambda\vartheta+i\omega(n+a-b+2l)/2)}
{\sinh(\lambda\vartheta-i\omega(n+a-b+2l)/2)}\ .
\label{gsa}
}
In the above $\Phi_{\bar ba}(\vartheta)$ is a scalar function which we
will not need to specify for the following argument.
Notice that, as required, $
\rho_{\bar ba}^j(\vartheta)=0$ for $j\not=0$
when $\vartheta=i\pi-i\pi(a-b)/N$ due to the factor with $l=0$ in the
numerator of \eqref{gsa} as long we fix 
\EQ{
N=n+2k\ .
}
This is an interesting result because it is consistent with intuition
from a completely different viewpoint. If we go back to
the Lagrangian formulation of the SSSG equations it is possible to 
proceed in an alternative way by treating the abelian subgroup of
$H=U(n)$ differently form the non-abelian part. For the latter, we can
only gauge the vector subgroup of $H_L\times H_R$. However, for the
$U(1)$ part we can choose to gauge the axial or the vector subgroup. This
gives a different formulation of the SSSG theory which is related by
T-duality to the ``usual'' formulation \cite{Miramontes:2004dr}. It is thought that T duality
is an exact quantum equivalence between theories. What is interesting is that in
this alternative theory there is a genuine $U(1)$ symmetry which is
not broken by the vacuum corresponding to abelian vector
transformations $\gamma\to U\gamma U^{-1}$, $U\in U(1)$.
In this formulation the soliton charge $q$ is a genuine Noether charge and we
may apply the Bohr-Sommerfeld quantization rule. This gives the
condition that $q=\pi a/2k$, for $a\in{\Bbb Z}$. If T-duality is
indeed an exact equivalence then this quantization of the charge $q$ is
consistent with the semi-classical limit of $q=\pi a/N$ with $N=n+2k$. 

Returning the kink/anti-kink S-matrix, we can repeat the analysis with
$a<b$, for which
\EQ{
R_{\bar ba}(\vartheta)=\Phi_{\bar b a}(\vartheta)
\sum_{j=0}^{a}(-1)^j\rho_{\bar ba}^j(\vartheta)
{\Bbb P}_{[b-a+2j,(b-a+j)^{n-2}]}\ ,
}
with 
\EQ{
\rho_{\bar ba}^j(\vartheta)=
\prod_{l=0}^{j-1}\frac{\sinh(\lambda\vartheta+i\omega(n+b-a+2l)/2)}
{\sinh(\lambda\vartheta-i\omega(n+b-a+2l)/2)}\ .
\label{gsa2}
}
Once again, as required, $
\rho_{\bar ba}^j(\vartheta)=0$ for $j\not=0$
when $\vartheta=i\pi-i\pi(b-a)/N$ due to the factor with $l=0$ in the
numerator of \eqref{gsa2}.

Crossing leads to a non-trivial equation for the scalar factor
$Y_{11}(\vartheta)$ which can be viewed as the unitarity constraint
for $S_{\bar b a}(\vartheta)$. It can be shown \cite{Hollowood:1993fj}
that this leads to the requirement
\EQ{
Y_{11}(i\pi-\vartheta)Y_{11}(i\pi+\vartheta)=
\frac{\sin(\omega+\pi\lambda-i\lambda\vartheta)
\sin(\omega+\pi\lambda+i\lambda\theta)}{\sin(\pi\lambda-i\lambda\vartheta)
\sin(\pi\lambda+i\lambda\theta)}\ .
\label{nn2}
}
The ``minimal'' solution---having no poles or zeros on the physical
strip---can be written most succinctly as a integral \cite{Hollowood:1993fj},
\EQ{
Y_{11}(\vartheta)&=
\exp\left[2i\int_0^\infty\frac{dt}t\,
\frac{\sin((n+2k)\vartheta t)\sinh((k+1)\pi t)\sinh(\pi t)}
{\sinh((n+k)\pi t)\sinh((n+2k)\pi t)}\right]\ .
}

{\sl The fusing rules}

The fusing rules summarize the direct channel bound states. In the
present theory they are
\EQ{
[a]\circ[b]&=\begin{cases}[a+b] &a+b\leq k\\ 0 & a+b>k\end{cases}\\
[a]\circ[b^{n-1}]&=\begin{cases}[a-b]&a>b\\ [(b-a)^{n-1}]&a<b\
  .\end{cases}
\label{swe}
}
These are a subset of the fusing rules of the minimal $A^{(1)}_{N-1}$ S-matrix.
We have shown that the simple poles in the S-matrix can all be
accounted for as bound-state poles in the either the direct or crossed
channels. Notice that the solution of the 
bootstrap is much simpler than the one considered in
\cite{Hollowood:1993fj} for which the kinks transformed in the
anti-symmetric representations. The reason being that the bootstrap
for the present case does not ``bite its own tail''
because crossing symmetry is much easier to implement 
arising from the fact that for the anti-symmetric
representations the anti-kinks arise as bound states of the kinks, and
hence non-trivial consistency conditions arise,
whereas for the symmetric representations they do not.

The $S$-matrix elements \eqref{hui} also have a series of double
poles. These will be interpreted exactly as for the $A_{N-1}$ minimal
S-matrix in terms of anomalous thresholds via the Coleman-Thun
mechanism \cite{Coleman:1978kk}.

{\sl The quantum group symmetry}

The $S$-matrix that we have constructed has an underlying quantum
group structure. In fact, the appropriate algebraic context is the
quantum loop group $U_q(SU(n)^{(1)})$ with $e^{\lambda\vartheta}$
  playing the r\^ole of the loop variable. 
Just like the ordinary group, the quantum group
$U_q(SU(n)$ can be generated by the Chevalley generators
$\{e_i,f_i,h_i\}$, $i=1,\ldots,n-1$ associated to the
simple roots. The affine quantum group involves 
adding in the generators for the highest root $e_0$ and $f_0$ with
powers of the loop variable.
The action of these generators on
the basic representations $[1]$ and $[1^{n-1}]$ is identical to the
ordinary group. What distinguishes a quantum group is how the
generators act on a tensor product. This describes the quantum group 
as a Hopf algebra with a co-product.
In contrast to the ordinary group,
on a tensor product $V\times U$ there is a non-trivial co-product
\EQ{
\Delta(h_i)&=h_i\otimes 1+1\otimes h_i\ ,\\
\Delta(e_i)&=e_i\otimes q^{-h_i}+q^{h_i}\otimes e_i\ ,\\
\Delta(f_i)&=f_i\otimes q^{-h_i}+q^{h_i}\otimes f_i\ .
}
The normal action is recovered in the limit $q\to1$. The $S$-matrix is
invariant under this action
\EQ{
\Delta(a)S(\vartheta)=S(\vartheta)\Delta(a)\ .
}

In addition, the $S$-matrix is invariant under a rapidity-dependent
symmetry which manifests the fact that it is actually invariant under the affine
symmetry $U_q(SU(n)^{(1)})$. 
Let $(e_0,f_0)$ be the raising and lowering operators
associated to the highest root and $h_0=-\sum_{i=1}^{n-1}h_i$. The
S-matrix also commutes with the action
\EQ{
\Big(e^{\lambda\vartheta_2}e_0\otimes q^{-h_0/2}&+
e^{\lambda\vartheta_1}q^{h_0/2}
\otimes e_0\Big)S(\vartheta)\\ & =S(\vartheta)
\Big(e^{\lambda\vartheta_1}e_0\otimes
q^{-h_0/2}+e^{\lambda\vartheta_2}q^{h_0/2}\otimes e_0\Big)\ ,
}
with a similar relation for $f_0$ with $e^{\lambda\vartheta_{1,2}}\to
e^{-\lambda\vartheta_{1,2}}$.

Notice that the action of the Cartan generators of the quantum group
is identical to the ordinary group and so the $S$-matrix has a
conventional $U(1)^n$ symmetry which is interpreted as a conserved
vector-valued topological charge.

\section{The Semi-Classical Limit}
\label{Semiclassical}

The scattering of solitons (or magnons) in an integrable field theory
has a very characteristic feature, the individual momenta, or
rapidities, of the solitons are conserved, however, a given soliton
can experience a rapidity-dependent time delay. 
The semi-classical limit relates this time delay
directly to the phase of the S-matrix and this provides a very 
stringent test of the S-matrix hypothesis. In the present case, the
semi-classical limit involves the level $k\to\infty$
and, in this limit, the phase shift $\delta$, defined by
$S=e^{2i\delta}$, is related to the classical time-delay $\Delta t(E)$ 
for two soliton scattering via the
WKB formula derived by Jackiw and Woo \cite{Jackiw:1975im}
\EQ{
\delta=\frac{n_B\pi}2+
\frac12\int_{E_\text{Th}}^E dE'\,\Delta t(E')\ ,
\label{del}
}
where $E=m_1\cosh\vartheta_1+m_2\cosh\vartheta_2$ is the 
energy in the COM frame and $E_\text{Th}$
is the threshold energy $E_\text{Th}=m_1+m_2$. The integer $n_B$ is the number of
bound states below threshold which will be 0 in the present context.
In this section we will use this formula
to test our S-matrix hypothesis. Note that the leading term of $\delta$
in the semi-classical limit scales like $k$ with corrections of order
$k^{-j}$, $j=0,1\ldots$. In particular, the constant term in
\eqref{del} only plays a role at leading order if the number of bound
states scales like $k$ which does not happen for the S-matrix in question.

\subsection{The classical time delay}

In order to calculate the time delay experienced by a soliton as it
scatters with another soliton, we need to specify the soliton's
space-time position in terms of the dressing data. The key quantity is 
\EQ{
\beta=\BF^\dagger\BF=\Bvarpi^\dagger\Psi_0(\xi)^{-1}\Psi_0(\xi^*)\Bvarpi\ ,
}
which for a soliton in isolation is
\EQ{
\beta=2e^{4F(x,t)}+1
}
where 
\EQ{
F(x,t)=\frac{\mu\sin q}2\big(
x\cosh\vartheta-t\sinh\vartheta\big)\ .
}
The spacetime position of the soliton can be identified with
the place where $F(x,t)=-\frac14\log2$, {\it
  i.e.\/}~$x=t\tanh\vartheta+\text{const}$.

The dressing transformation makes it simple to extract the classical time delay experienced by a magnon/soliton as it
scatters with another magnon/soliton. The idea is to focus on the
spacetime position of soliton 2 and think 
of it as dressed by soliton 1. As for the soliton in isolation,
the position of soliton 2 is encoded in the quantity
\EQ{
\beta^{(2)}=\Bvarpi^{(2)\dagger}
\Psi^{(1)}(\xi_2)^{-1}\Psi^{(1)}(\xi_2^*)\Bvarpi^{(2)}
}
where now we have the dressed quantity
\EQ{
\Psi^{(1)}(\lambda)=\chi^{(1)}(\lambda)\Psi_0(\lambda)\ .
}

In order to calculate the time delay, or spacetime shift, we
then need to take the limits of $\chi^{(1)}(\lambda)$ as
$x\to\pm\infty$. This follows from
\EQ{
\chi(\lambda)\xrightarrow[x\to\infty]{}
{\bf1}+\frac12\frac{\xi-\xi^*}{\lambda-\xi}
\left(\begin{array}{cc|c}1 & - i & \B0\\ 
+ i &1 & \B0 \\ \hline
\B0 & \B0  &
\B0\end{array}\right)+\frac12\frac{\xi^{-1}-\xi^{*-1}}{\lambda-\xi^{-1}}
\left(\begin{array}{cc|c}1 & + i & \B0
\\
- i &1 & \B0 \\ \hline
\B0 & \B0  &
\B0\end{array}\right)
}
and
\EQ{
\chi(\lambda)\xrightarrow[x\to-\infty]{}&
{\bf1}+A^{-1}\Big[\frac1{\lambda-\xi}\Big(\frac\xi{|\xi|^2-1}-
\frac{|\xi|^2}{\xi-\xi^*}\Big)\\ &+\frac1{\lambda-\xi^{-1}}
\Big(\frac1{\xi-\xi^*}-
\frac{\xi^*}{|\xi|^2-1}\Big)\Big]\left(\begin{array}{c|c}\B0 & \B0
\\\hline
\B0 &\BOmega\BOmega^\dagger
\end{array}\right)\ ,
}
with
\EQ{
A=\frac{|\xi|^2}{(|\xi|^2-1)^2}-\frac{|\xi|^2}{(\xi-\xi^*)^2}\ .
}

From these we deduce that soliton 2 has $F_2(x,t)$ shifted by 
\EQ{
\Delta F=\log\left[\left|\frac{\xi_1-\xi_2}{\xi_1-\xi_2^*}\right|^4
\left|\frac{1-\xi_1\xi_2}{1-\xi_1\xi_2^*}\right|^2\cos^2\Theta+
\left|\frac{\xi_1-\xi_2}{\xi_1-\xi_2^*}\right|^2\sin^2\Theta\right]\ ,
\label{delf}
}
where we have taken $|\BOmega^{(2)*}\cdot\BOmega^{(1)}|=\cos\Theta$. 
The corresponding time delay is
\EQ{
\Delta t=\frac{\Delta F}{\mu|\sin q_2|\sinh\vartheta_2}\ .
}
The result matches that in \cite{Hatsuda:2009pc}
(see also \cite{Kalousios:2010ne})\footnote{The consistency of the classical time delays with the $\text{AdS}_4/\text{CFT}_3$ magnon $S$-matrix conjectured by Ahn and Nepomechie~\cite{Ahn:2008aa} has been checked in~\cite{Hatsuda:2009pc}.}. 
In the COM frame $|\sin q_1|\sinh\vartheta_1=-|\sin
q_2|\sinh\vartheta_2$ and so we can write the formula for the phase
shift in a manifestly relativistic way as
\EQ{
\delta=\frac{n_B\pi}2+
\frac{k}{\pi}\int_0^{\vartheta} d\vartheta'\,\Delta F(\vartheta')\ ,
}
where $\Delta F$ in \eqref{delf} can be written in terms of the
rapidity difference
\EQ{
\Delta F(\vartheta)&=\log\left[\left|\frac{
\sinh(\frac\vartheta2-i\frac{q_1-q_2}2)}
{\sinh(\frac\vartheta2-i\frac{q_1+q_2}2)}
\right|^4\left|\frac{
\cosh(\frac\vartheta2-i\frac{q_1-q_2}2)}
{\cosh(\frac\vartheta2-i\frac{q_1+q_2}2)}
\right|^2\cos^2\Theta\right.\\ &~~~~~~~~~~~~~~+
\left.\left|\frac{
\sinh(\frac\vartheta2-i\frac{q_1-q_2}2)}
{\sinh(\frac\vartheta2-i\frac{q_1+q_2}2)}
\right|^2\sin^2\Theta\right]\ .
\label{delft}
}

\subsection{Taking the semi-classical limit of the S-matrix}
\label{SemiC}

In order to take the semi-classical limit of our S-matrix we need to
specify carefully which particular quantum states can be discussed. 
The states which have a good semi-classical limit are those with a
fixed charge $q$ as $k\to\infty$. This means that for states in
representation $[a]$, $a$ must also $\to\infty$ with $a/k$ fixed. In
other words the good semi-classical states are in large symmetric
representations. Actually this is very natural, the classical solitons
have an internal collective coordinate $\BOmega$ valued in
$\CP^{n-1}$, since the phase of $\BOmega$ is irrelevant. To
semi-classically quantize this degree-of-freedom, one lets the
collective coodinates become time-dependent and plugs this into the
action to yield an effective quantum-mechanical action. This action
turns out to be first order in the time derivatives and the resulting
quantization is not conventional.
Rather, as we shall argue, the classical moduli space
itself should be thought of as a symplectic manifold in its own right
and quantized accordingly. The classical moduli space then emerges 
from the semi-classical limit of a ``fuzzy'' geometry.\footnote{This
  will be explained in more detail in the companion paper \cite{us2}.}

This classical moduli
space has a description in terms of an adjoint orbit of $SU(n)$ 
since we can rotate $\BOmega$ by means of a global $H$ transformation on the
soliton solution: $\gamma\to U\gamma U^{-1}$ implies $\BOmega\to
U\BOmega$.
If we describe the classical moduli space in terms of $h_\BOmega$, the infinitesimal generator of $H=U(n)\subset SU(n+2)$ defined in~\ref{defh}, the adjoint orbit is then of the form
$h_\BOmega = U\text{diag}(1,1,-2,0,\ldots,0)U^{-1}$. More intrinsically, if 
we project out the
part of $h_\BOmega$ lying inside the Lie algebra of $SU(n)$ (and
re-scale appropriately) then the 
adjoint orbit is the one through the $SU(n)$ Lie algebra element 
$\text{diag}(-n+1,1,\ldots,1)$ which is another way to define
the projective space $\CP^{n-1}$. 

Quantization of the collective coordinate moduli space spanned by $\BOmega$ 
involves pulling back the symplectic form of the WZW model to
classical moduli space and
leads to the quantization of the adjoint orbit (or co-adjoint orbit
since these are the same for semi-simple Lie groups). In the case to
hand, the quantization on the co-adjoint orbit 
is known as the fuzzy $\CP^{n-1}$ \cite{Balachandran:2005ew}. The
coordinates on the space are to be thought of as quantum operators on a
Hilbert space, and hence are non-commuting.
The resulting quantum Hilbert space of 
states are the symmetric representations of 
$SU(n)$ and as the dimension of the representations becomes large 
the fuzzy $\CP^{n-1}$ becomes a closer approximation to the ``ordinary''
space.
To see this, let $|\Be_i\rangle$, $i=1,\ldots,n$, be a basis
for the $n$-dimensional module of $SU(n)$. So $\Be_i$ is the soliton
with topological charge $\Be_i$. Consider the
special states $||\BOmega,a\rangle\rangle=
\big(\BOmega_i|\Be_i\rangle\big)^{\otimes a}$. These
states lie in the module $[a]$ and have an inner-product
\EQ{
\langle\langle\BOmega,a||\BOmega',a\rangle\rangle=\big(\BOmega^*\cdot
\BOmega'\big)^a
}
which goes to zero as $a\to\infty$ if
$\BOmega'\neq\BOmega$. Consequently, as $a\to\infty$, the quantum state
$||\BOmega,a\rangle\rangle$ is a quasi-classical state (or coherent
state) which
approximates the classical configuration with collective coordinate
$\BOmega$. Actually, since the symmetry in the present context is a
quantum group symmetry we expect that the quantization of the solitons
involves a fuzzy $\CP^{n-1}$ with a $q$-deformation
\cite{Grosse:2000gd,Pawelczyk:2002kd}. Notice
that the $q$ deformation will only become apparent for states where
$a$ is of order $k$.

In the following, we shall focus on the particular states
$||\Be_i,a\rangle\rangle$ for simplicity. These states are associated
to the special set of classical solitons with $\BOmega=\Be_i$
described previously which have a topological charge that is aligned
with the choice of gauge. The S-matrix elements for
these states follow in a simple way by fusing the basic elements for
the solitons with charges $\Be_i$. These elements are
given by \eqref{SMAT}. We define the
scattering of $|\Be_i\rangle$ with $|\Be_i\rangle$, that is two kinks with
topological charge $\Be_i$, as
\EQ{
S_1(\vartheta)&=X_{11}(\vartheta)Y_{11}(\vartheta)
\rc_{11}\left(\left.\ARR{\Ba&\Ba+\Be_i\\
      \Ba+\Be_i&\Ba+2\Be_i}\right\vert\vartheta\right)\\
&=Y_{11}(\vartheta)
\frac{\sinh(\frac{\lambda\vartheta}2
+\frac{i\omega}2) \cosh(\frac{\lambda\vartheta}2+\frac{i\omega}4)}
{\sinh(\frac{\lambda\vartheta}2-\frac{i\omega}2)
  \cosh(\frac{\lambda\vartheta}2-\frac{i\omega}4)}\ .
}
The scattering of two kinks with topological charge $\Be_i$ and
$\Be_j$, $i\neq j$ has both a transition amplitude and a reflection amplitude. In order to
take the semi-classical limit we need only consider the transition
amplitude which we define as
\EQ{
S_2(\vartheta)&=X_{11}(\vartheta) Y_{11}(\vartheta)
\rc_{11}\left(\left.\ARR{\Ba&\Ba+\Be_j\\ \Ba+\Be_i&\Ba+\Be_i+\Be_j}\right\vert\vartheta\right)\\
&=X_{11}(\vartheta) Y_{11}(\vartheta)
 {\sinh(\lambda\vartheta)\over\sinh(\lambda\vartheta+i\omega)}{\cal
   Z}\\
&=Y_{11}(\vartheta)\frac{\sinh(\frac{\lambda\vartheta}2
+\frac{i\omega}2) \cosh(\frac{\lambda\vartheta}2+\frac{i\omega}4) \sinh(\lambda\vartheta)}
{\sinh(\frac{\lambda\vartheta}2-\frac{i\omega}2)
  \cosh(\frac{\lambda\vartheta}2-\frac{i\omega}4)\sinh(\lambda\vartheta
+i\omega)}{\cal Z}\ ,
}
where ${\cal Z}$ is the square root factor in \eqref{hj2}
that depends on the vacuum state. Since this factor is a real number
and we are only interested in the phase of the S-matrix, it
will play no r\^ole in what follows. We have written the factor
$Y_{11}(\vartheta)$ explcitly since as $k\to\infty$, $\log
Y_{11}(\vartheta)$ is order $k^{-1}$ and therefore is subleading and
can be ignored.

We can then calculate the scattering of the quasi-classical state
$||\Be_i,a\rangle\rangle$
with $||\Be_j,b\rangle\rangle$ by applying the bootstrap
equations. For $i=j$, and taking $a\geq b$\footnote{In the following
  we do not indicate the $Y_{ab}(\vartheta)$ factors because they are subleading.}
\EQ{
S_{ia,ib}(\vartheta) &=\prod_{j=0}^{a-1}\prod_{l=0}^{b-1}
S_{1}\Big(\vartheta+\frac{i\pi(a+b-2j-2l-2)}N\Big)\\
&=\prod_{j=1}^b\frac{(a+b-2j+2) _\vartheta (a+b-2j) _\vartheta[a+b-2j+1]_\vartheta}
{(2j-a-b-2) _\vartheta (2j-a-b) _\vartheta[2j-a-b-1]_\vartheta}\ ,
\label{zs1}
}
while for $i\neq j$ 
\EQ{
S_{ia,jb}(\vartheta) &=\prod_{j=0}^{a-1}\prod_{l=0}^{b-1}
S_2\Big(\vartheta+\frac{i\pi(a+b-2j-2l-2)}N\Big)\\
&={\cal Z}_{ia,jb}
\prod_{j=1}^b\frac{(a+b-2j) _\vartheta[a+b-2j+1]_\vartheta[2j-a-b]_\vartheta}
{(2j-a-b-2) _\vartheta[2j-a-b-1]_\vartheta[a+b-2j+2]_\vartheta}\ .
\label{zs2}
}
where the functions $(x) _\vartheta$ and $[x]_\vartheta$ are defined in \eqref{dem} and
${\cal Z}_{ia,jb}$ is a real-valued vacuum-dependent factor.

We now have the S-matrix elements in a form that is suitable for
taking the semi-classical limit. As $k\to\infty$, keeping the charges
$q_1=a\omega/2\lambda$ and $q_2=b\omega/2\lambda$ fixed, the 
products over $j$ in \eqref{zs1} and \eqref{zs2} can be expressed as
an integral over a continuous variable:
\EQ{
\prod_{j=1}^{b}f\Big(\frac{\lambda\vartheta}2\pm\frac{i\omega(a+b-2j+l)}4\Big)
\longrightarrow \exp\left[\frac{4k}\pi\int_{q_1-q_2}^{q_1+q_2}d\eta\,
\log f\Big(\frac{\vartheta}2\pm\frac{i\eta}2\Big)\right]\ .
}
In the above $l$ is arbitrary as long as it is fixed as $k\to\infty$.
We now write the integral over $\eta$ as an integral over $\vartheta$
using the identity
\EQ{
i\int_{q_1-q_2}^{q_1+q_2}d\eta\,\log\left[\frac{
f(\frac\vartheta2-\frac{i\eta}2)f(\frac{i\eta}2)}
{f(\frac\vartheta2+\frac{i\eta}2)f(-\frac{i\eta}2)}\right]
=\int_0^{\vartheta}d\vartheta'\,\log\left|\frac{
f(\frac{\vartheta'}2-\frac{i(q_1-q_2)}2)}
{f(\frac{\vartheta'}2-\frac{i(q_1+q_2)}2)}
\right|
}

One can then check that in the semi-classical limit
\EQ{
\IM\log S_{ia,ib}(\vartheta)=\frac{2k}\pi\int_0^\vartheta
  d\vartheta'\,\Delta F(\vartheta')\Big|_{\Theta=0}
}
and 
\EQ{
\IM\log S_{ia,jb}(\vartheta)=\frac{2k}\pi\int_0^\vartheta
  d\vartheta'\,\Delta F(\vartheta')\Big|_{\Theta=\frac\pi2}\ .
}
This completes our check of the S-matrix via the semi-classical limit.

\section{The Case $\CP^2$}
\label{CP2}

Strictly speaking our analysis only applies to the case where the
group $H$ is non-abelian and so this excludes the case $\CP^2$ for
which $G/H=U(2)/U(1)$. In this section we consider this case and find
some similarities but also some differences. Importantly, in this case
we are able to test the S-matrix against both the semi-classical limit
but also against the Thermodynamic Bethe Ansatz (TBA) via which one
can calculate the central charge of the UV CFT: the $U(2)_k/U(1)$ gauged WZW model.

Our conjectured S-matrix for $\CP^2$ is based on a spectrum of states which 
matches \eqref{AMAS}, so that $N=2k+1$. 
The S-matrix elements are then conjectured to be
\EQ{
S_{ab}(\vartheta)=\widetilde\eta(a,b) X_{ab}(\vartheta)
}
$\widetilde\eta(a,b)$ is a constant phase that is needed to satisfy crossing and
bootstrap (we will fix it below). In the above, $X_{ab}(\vartheta)$ is
defined as in \eqref{uy3} but with
$\lambda=1$ and $\omega=2\pi/N$ where we interpret the
labels $a,b$ as defined modulo $N=2k+1$. The S-matrix has a pole structure
on the physical strip that matches the 
minimal S-matrix associated to $A_{2k}^{(1)}$. This means that the
fusing rules allow more bound states than \eqref{swe}, with $a\circ
b=a+b\text{ mod }N$. Moreover, the particle labelled by $\overline{a}=N-a$ is identified with the anti-particle of the particle labelled by $a$. In fact if we write the (diagonal) S-matrix as 
\EQ{
S_{ab}(\vartheta)=\widetilde\eta(a,b) S^{\text{min}}_{ab}(\vartheta)
S^{\text{CDD}}_{ab}(\vartheta),
}
then $S^{\text{min}}_{ab}$ is the minimal S-matrix associated to $A_{2k}^{(1)}$ and
the CDD part is related to the CCD part of the S-matrix of the
homogeneous sine-Gordon theory 
models at level $N$ (see \cite{Dorey:2004qc}, section 4): 
\EQ{
S^{\text{CDD}}_{ab}(\vartheta)=\Big[S^{\text{F}}_{\overline 
ab}(\vartheta)\Big]^{-1}\ ,
}
where
\EQ{
S^{\text{F}}_{ab}(\vartheta)=\prod_{j=1}^{\text{min}(a,b)}
\frac{(a+b-2j+1) _\vartheta}{(2j-a-b-1) _\vartheta}\ .
}
Then, the resulting set of TBA equations is
\EQ{
\varepsilon_a(\vartheta)= \nu_a(\vartheta) -\sum_{b=1}^{N-1} \big(\phi_{ab} +\psi_{\overline {a}b}\big)\ast L_b(\vartheta),
\label{TBAnew}
}
with 
\SP{
&\nu_a =m_a r\cosh\vartheta,\qquad
L_a = \log \big(1+e^{-\varepsilon_a}\big),\\[5pt]
&\phi_{ab}=-i\frac{d}{d\vartheta} S_{ab}^\text{min}(\vartheta),\qquad
\psi_{ab}=+i\frac{d}{d\vartheta} S_{ab}^{\text{F}}(\vartheta),
}
whose scaling function is
\EQ{
c(r)=\frac{3}{\pi^2} \sum_{a=1}^{N-1} \int_{-\infty}^{+\infty} d\vartheta\> \nu_a(\vartheta) L_a(\vartheta).
}
Taking into account that $\nu_a=\nu_{\overline{a}}$, $\phi_{ab}=\phi_{\overline{a}\overline{b}}$ and $\psi_{ab}=\psi_{\overline{a}\overline{b}}$, it follows that $\varepsilon_{a}=\varepsilon_{\overline{a}}$, and the system of equations~\eqref{TBAnew} can be written in the equivalent way
\EQ{
\varepsilon_a(\vartheta)= \nu_a(\vartheta) -\sum_{b=1}^{N-1} \big(\phi_{ab} +\psi_{ab}\big)\ast L_b(\vartheta).
\label{TBAnewB}
}

Taking advantage of the fact that $\phi_{ab}$ and $\psi_{ab}$ are the kernels that enter the TBA equations of the HSG models, we can relate our set of TBA equations to those of the $SU(3)_N$ HSG model:
\SP{
&
\varepsilon_a^1(\vartheta)= \nu_a^1(\vartheta) -\sum_{b=1}^{N-1} \big(\phi_{ab}\ast L_b^1(\vartheta) +\psi_{ab}\ast L_b^2(\vartheta-\sigma_{21})\big)\\[5pt]
&
\varepsilon_a^2(\vartheta)= \nu_a^2(\vartheta) -\sum_{b=1}^{N-1} \big(\phi_{ab}\ast L_b^2(\vartheta) +\psi_{ab}\ast L_b^1(\vartheta-\sigma_{12})\big),
}
where $\nu_a^i =M_i m_a r\cosh\vartheta$. For any non-vanishing value of the mass scales $M_1$ and $M_2$, and any value of the resonance parameters $\sigma_{12}=-\sigma_{21}$, it was shown in~\cite{CastroAlvaredo:1999em} that
\EQ{
c(r)=\frac{3}{\pi^2} \sum_{i=1}^2\sum_{a=1}^{N-1} \int_{-\infty}^{+\infty} d\vartheta \nu_a^i(\vartheta) L_a^i(\vartheta) \longrightarrow  \frac{6(N-1)}{N+3}
}
when $r\rightarrow 0$, which is the central charge of the $SU(3)_N/U(1)^2$ coset CFT.

Let us consider the particular choice of parameters $M_1=M_2=1$, and $\sigma_{12}=0$. Then the $SU(3)_N$ TBA equations simplify to
\SP{
&
\varepsilon_a^1(\vartheta)= \nu_a(\vartheta) -\sum_{b=1}^{N-1} \big(\phi_{ab}\ast L_b^1(\vartheta) +\psi_{ab}\ast L_b^2(\vartheta)\big)\\[5pt]
&
\varepsilon_a^2(\vartheta)= \nu_a(\vartheta) -\sum_{b=1}^{N-1} \big(\phi_{ab}\ast L_b^2(\vartheta) +\psi_{ab}\ast L_b^1(\vartheta)\big).
}
Obviously, $\varepsilon_a^1(\vartheta)=\varepsilon_a^2(\vartheta)$, and we obtain two identical copies of the system~\cite{CastroAlvaredo:1999em}
\EQ{
\varepsilon_a(\vartheta)= \nu_a(\vartheta) -\sum_{b=1}^{N-1} \big(\phi_{ab}\ast L_b(\vartheta) +\psi_{ab}\ast L_b(\vartheta)\big),
}
which is just~\eqref{TBAnewB}. Therefore,
\EQ{
c(r)=\frac{3}{\pi^2} \sum_{a=1}^{N-1} \int_{-\infty}^{+\infty} d\vartheta \nu_a(\vartheta) L_a(\vartheta) \longrightarrow  \frac{3(N-1)}{N+3};
}
namely, one half the UV central charge of the $SU(3)_N/U(1)^2$ coset CFT.

Our hypothesis is that $N=2k+1$ and so
\EQ{
c_{CFT} = \frac{3k}{k+2}.
}
which is precisely the central charge of the $U(2)_k/U(1)$ coset CFT.
Finally, let us fix the overall phases in $S_{ab}$. As pointed out in~\cite{Miramontes:1999hx},
\EQ{
S^{\text{F}}_{ab}(i\pi-\vartheta)= (-1)^a S^{\text{F}}_{\overline{b}a}(\vartheta).
}
Therefore, since $N=2k+1$ is odd, it is not difficult to check that
\EQ{
\widetilde\eta(a,b)=(-1)^{ab}
}
ensures that the S-matrix satisfies the usual crossing and bootstrap relations. Obviously, the overall constant phase plays no r\^ole in the TBA equations.

\section{Discussion}\label{Comparison}

The purpose of this paper has been to begin the programme of solving
the SSSG theories at the quantum level with the goal of seeing to what extent the
spectrum and S-matrix of these relativistic theories is related to
their non-relativistic Pohlmeyer cousins that describe giant magnons
in string theory. It is not expected that
there will be an exact equivalence of any kind unless the full problem
for the
supergroup symmetric space models is considered. However, we have seen
that the soliton
theory does have certain things common with the magnon theory in that
states transform in symmetric representations of the symmetry. 
However, even at the classical level, there is a non-trivial
rapidity dependent mapping between the charges of the magnons and
solitons. In addition, in the
soliton case, the symmetry is a affine quantum group symmetry,
whereas in the magnon
case it is a ``normal'' symmetry. Both solitons and magnons come in a
tower of states, however, for the solitons the tower is truncated by
the quantum group structure. 
It is clear that if there is some some
kind of equivalence for the cases involved in the AdS/CFT
correspondence then we can expect some surprises for the supergroup
extensions of the SSSG theories.

It is interesting to consider how the quantum solution of the deformed
WZW model relates to the field theory in the classical
limit. Classically, the theory has a degenerate vacuum that can be
identified with the Cartan torus of $H$. In the quantum theory, the
set of vacua is the discrete set $\Lambda^*(k)$. However, as
$k\to\infty$ there is an obvious sense in which this discrete set can be
described by a continuum taking values in the Cartan torus. The solitons
in the quantum theory are kinks whose topological charge takes values
in the set of weights of the symmetric representations, which again as
$k\to\infty$ become arbitrary vectors in the Cartan space. In fact, we
have already mentioned that the internal
$\CP^{n-1}$ moduli space of the classical soliton can be viewed as
becoming  
a $q$-deformed fuzzy $\CP^{n-1}$ in the semi-classical
approximation. It is important to emphasize that the S-matrix we have
written down is subject to the CDD ambiguities and the semi-classical
limit only partially constrains these. In order to pin them down
definitively, one should perform a TBA analysis for all the
$\CP^{n+1}$ cases; a task that will be pursued  elsewhere.

It would be interesting to compare our S-matrix with the approach to
quantizing the deformed WZW model adopted
in \cite{Hoare:2009fs}. In that reference the approach taken is
essentially perturbative, in that fields are expanded around the vacuum in a
particular gauge which involves setting ${\cal A}_+=0$ and then
integrating out ${\cal A}_-$ to give a non-local form of the
action. This non-local action then has an equivalent local form and
the tree-level $S$-matrix can be computed. In our approach, we expect
that the perturbative excitations of the theory correspond to states
with lowest $U(1)$ charge, $q=\pm\pi/(2k+n)$. Indeed, in the semi-classical
limit, these states have a perturbative mass $M=\mu$. In our approach
these states are kinks but with vanishing small topological charge in
the semi-classical  limit.

This paper only presents the first step in understanding the SSSG
theories at the quantum level. Generalizations to other symmetric
spaces are currently under way. A particularly important class of
examples are the symmetric spaces
$F=SO(n+2)/SO(n+1)\simeq S^{n+1}$, for which the associated SSSG
equations involve the WZW theory for coset
$G/H=SO(n+1)/SO(n)\simeq S^n$. The solitons in this case, have a classical
moduli space which has an adjoint orbit of $SO(n)$ identified with the
real oriented
Grassmannian $SO(n)/SO(2)\times SO(n-2)$. The quantum
states in this case correspond to symmetric representations of
$SO(n)$. The S-matrices for these theories will be described in a
companion paper \cite{us2}.

\acknowledgments

\noindent
JLM acknowledges the support of MICINN (Spain) 
and FEDER (FPA2008-01838 and 
FPA2008-01177), Xunta de Galicia (INCITE09.296.035PR), and the 
Spanish Consolider-Ingenio 2010
Programme CPAN (CSD2007-00042).

\noindent TJH would like to thank Nick Dorey for useful conversations and the 
organizers of the conference ``16 Supersymmetries'' at City University
London in May at which these results were first presented. TJH would
also like to acknowledge the support of STFC grant
ST/G000506/1.

We would both like to thank Arkady Tseytlin for discussions and
comments on a draft of this paper.

\end{document}